\documentclass{elsarticle}

\usepackage{amssymb}
\usepackage{amsmath}
\usepackage[table]{xcolor}
\usepackage{enumitem}
\usepackage{tcolorbox}
\usepackage{adjustbox}
\usepackage{hyperref}
\usepackage{float} 
\newlist{questions}{enumerate}{1}
\setlist[questions]{label=\textbf{RQ\arabic*:}}

\newcommand{\header}[1]{\medskip\noindent\textbf{#1}}

\newcommand{\rqi}{Does GPT-assistance during the PR process reduce the resolution time?}
\newcommand{\rqii}{How does GPT assistance impact various phases of the PR process?}
\newcommand{\rqiii}{Why do contributors and reviewers use GPT during the PR process?}

\journal{Information and Software Technology}

\begin{document}

\begin{frontmatter}

\title{The Impact of Large Language Models (LLMs) on Code Review Process}

\author[label1]{Antonio Collante\corref{cor1}}\ead{antonio.collante@mail.concordia.ca}
\author[label1]{Samuel Abedu}\ead{samuel.abedu@mail.concordia.ca}
\author[label1]{SayedHassan Khatoonabadi}\ead{sayedhassan.khatoonabadi@concordia.ca}
\author[label2]{Ahmad Abdellatif}\ead{ahmad.abdellatif@ucalgary.ca}
\author[label1]{Ebube Alor}\ead{ebubechukwu.alor@mail.concordia.ca}
\author[label1]{Emad Shihab}\ead{emad.shihab@concordia.ca}
\affiliation[label1]{organization={Concordia University},
             addressline={2155 Guy St},
             city={Montreal},
             postcode={H3H 2L9},
             state={Quebec},
             country={Canada}}
\affiliation[label2]{organization={University of Calgary},
             addressline={2500 University Dr NW},
             city={Calgary},
             postcode={T2N 1N4},
             state={Alberta},
             country={Canada}}
\cortext[cor1]{Corresponding author}

\begin{abstract}
Large language models (LLMs) have recently gained prominence in the field of software development, significantly boosting productivity and simplifying teamwork. Although prior studies have examined task-specific applications, the phase-specific effects of LLM assistance on the efficiency of code review processes remain underexplored. This research investigates the effect of GPT on GitHub pull request (PR) workflows, with a focus on reducing resolution time, optimizing phase-specific performance, and assisting developers. We curated a dataset of 25,473 PRs from 9,254 GitHub projects and identified GPT-assisted PRs using a semi-automated heuristic approach that combines keyword-based detection, regular expression filtering, and manual verification until achieving 95\% labeling accuracy. We then applied statistical modeling, including multiple linear regression and Mann–Whitney U test to evaluate differences between GPT-assisted and non-assisted PRs, both at the overall resolution level and across distinct review phases. Our research has revealed that early adoption of GPT as assistant can boost the effectiveness of the PR process, leading to considerable time savings at various stages. Our findings suggest that GPT-assisted PRs reduced median resolution time by more than 61\%. We discovered that utilizing GPT can reduce the review time by 66.7\%, and the waiting time before acceptance by 87.5\%. Analyzing a sample of 300 GPT-assisted PRs, we discovered that developers predominantly use GPT for code optimization (60\%), bug fixing (26\%), and documentation updates (12\%). Our findings highlight the potential of GPT to accelerate code reviews, improve code quality, and address persistent bottlenecks. This research sheds light on the impact of the GPT model on the code review process, offering actionable insights for software teams seeking to enhance workflows and promote seamless collaboration.
\end{abstract}

\begin{keyword}
LLMs \sep GPT \sep AI assistance \sep pull requests \sep pull request phases \sep code review process \sep developer productivity.
\end{keyword}

\end{frontmatter}

\section{Introduction}
\label{sec:introduction}

Modern software development is heavily based on collaborative workflows, with Pull Requests (PR) serving as a central mechanism to integrate code changes \cite{gousios2014exploratory, sadowski2018modern, bacchelli2013expectations}. A PR goes through multiple phases, including submission, review, change, and merging, each requiring coordination between contributors and reviewers \cite{gousios2014exploratory}. Effective PR management is crucial to maintaining software quality, but it remains a bottleneck in many projects due to long review times, inconsistent feedback, and inefficient communication \cite{zhang2022pull}. Developers often experience delays due to unclear review comments, excessive revisions, and prolonged waiting times before changes are accepted \cite{kudrjavets2022mining}.

Inefficiencies in the software development cycle, such as long review times, can lead to frustration among team members, decreased productivity, and even project delays \cite{baysal2016investigating}. As codebases become increasingly intricate, there is a growing need for sophisticated automation tools to facilitate the code review process and improve collaboration and teamwork \cite{zheng2024code}. In parallel, LLMs, such as Generative Pre-trained Transformers (GPTs) \cite{vaswani2017attention} have rapidly evolved, demonstrating strong capabilities in code generation, bug detection, and documentation assistance \cite{tufano2024unveiling, ebert2023generative}. These models are capable of suggesting enhancements to existing code, resolving unclear logical structures, and automating repetitive tasks, thereby becoming indispensable tools for software engineers.

Despite the increasing adoption of LLM-powered tools in development workflows, a systematic investigation into their impact on PR processes is lacking. Previous research on the use of LLM in the PR context \cite{tufano2024unveiling,grewal2024analyzing, chouchen2024so, rasheed2024ai} has focused mainly on task-specific applications, such as commit message generation \cite{xiao2024generative} and documentation support \cite{davila2024tales}, while neglecting to explore their phase-specific impact across the entire PR lifecycle. Furthermore, while LLM-generated suggestions can be beneficial, it is necessary to quantify their effectiveness in improving review speed and reducing PR merge time.

This study aims to fill this knowledge gap by empirically evaluating the impact of GPT on code review waiting times, with a particular focus on phase-specific effects. It examines their influence on resolution time, their role in different PR phases, and the reasons why contributors and reviewers adopt GPT as assistance. Given the proven effectiveness of GPT on a wide range of software engineering tasks, including code generation, debugging, and documentation, understanding how and to what extent GPT enhances the effectiveness of PRs in each phase is crucial to assess their usefulness in real-world operations. These insights can guide the design of more efficient GPT-supported code review workflows and inform best practices for integrating AI assistance into collaborative software development.

Toward this goal, we examine the PRs that software developers actively engage with GPT in conversational exchanges to seek recommendations, clarifications, or solutions related to code review tasks. We refer to these as \textbf{GPT-assisted PRs} throughout the paper. For example, in this GPT-assisted PR~\cite{pullrequest2059}, approximately 99\% of the code was generated by GPT-4, as the developer requested ChatGPT to enhance a Ruby plugin that enables Jekyll to fetch posts from an external RSS feed when rss\_url is specified. In contrast, \textbf{GPT non-assisted PRs} are those where developers either do not use GPT at all or use it strictly for implementation tasks (i.e., making an API call to a GPT model), without relying on it for code review insights or decision making. Our research investigates how GPT-assisted PRs influence the duration and quality of the code review process. For this purpose, we start by curating a dataset of 25,473 PRs from 9,254 projects on GitHub. Next, we extract six features (\textit{project\_age, no\_commits, files\_added, files\_deleted, files\_changed, time\_to\_merge}) to characterize projects, PRs, and code reviews. We then analyze how these factors influence the review and resolution times \cite{moreira2021factors}. Finally, we also investigate which development tasks involve the use of GPT-assisted PRs throughout the PR phases, aiming to identify where such assistance is most effectively integrated. In summary, our aim is to answer the following three research questions:

\begin{questions}
\item \textbf{\rqi} We aim to understand the potential of LLMs to expedite the resolution of PRs in the code review process. Specifically, we analyze their impact on the total time required to resolve a PR, from its submission to completion. \textit{Our results suggest a correlation between the use of GPT-assisted PRs and a decrease in merge time approximately from 23 to 9 hours}.

\item \textbf{\rqii} We aim to further explore the effect of GPT in various phases of the PR process, such as submission, evaluation, modification, and waiting time, to determine where its support is most advantageous. \textit{Our analysis shows that GPT-assisted PRs complete faster across key phases of the review process. Specifically, they take 1 hour at the Review phase, 3 hours at Waiting for Changes, and 1 hour at Change, compared to 3 hours, 24 hours, and 1 hour, respectively, for non-assisted PRs.} These results highlight the potential of GPT to streamline the PR workflow, especially by reducing waiting times.

\item \textbf{\rqiii} We aim to better understand how programmers use GPT to enhance code quality, fix errors, and improve documentation. By analyzing these application patterns, we uncover how developers integrate this technology into their daily workflows, highlighting its most potential benefits. \textit{Our findings show that developers devote a significant part of their review time to refining and correcting the source code. Specifically, 60.26\% of enhancement tasks occur during the Review phase, 42.31\% during Waiting for Changes, and 35.90\% during the Change phase.} This explains the role of GPT as assistance in enhancing efficiency for those time-consuming activities.
\end{questions}

Finally, our research provides recommendations for software developers, both maintainers and contributors, as well as researchers, aimed at optimizing and accelerating the PR review process. We argue that applying the GPT recommendations as a support tool can strengthen collaboration by allowing participants to better anticipate and resolve potential issues.

\header{Our Contributions.} In summary, we make the following contributions in this paper:
\begin{itemize}
    \item To the best of our knowledge, this is the first comprehensive analysis that examines the effectiveness of GPT during the PR review phases.
    \item We present empirical evidence on the impact of GPT on PR-based development phases and identify the types of software development activities where it is used the most frequently.
    \item To promote the reproducibility of our study and facilitate future research on this topic, we publicly share our scripts and dataset online at \url{https://github.com/acollant/GPT-Assistance-PR}.
\end{itemize}

\header{Paper Organization.} The remainder of the paper is organized as follows. We review related work in Section \ref{sec:relatedwork} and describe our study setup and dataset in Section \ref{sec:setup}.  We report our approach and findings for each research question in Section \ref{sec:results} and discuss the implications of our findings in Section \ref{sec:discussion}. Finally, we describe the limitations of our study in Section \ref{sec:threats} and conclude the paper in Section \ref{sec:conclusion}.
\section{Related Work}
\label{sec:relatedwork}
In the following, we first review research on the challenges in code reviews and collaboration before moving on to an overview of factors affecting PR lifetime and latency. Finally, we review studies on the application of LLMs in code reviews.

\header{Challenges in Modern Code Reviews (MCR) and Collaboration.} 
Code reviews are a critical aspect of modern software development, ensuring code quality, maintainability, and adherence to best practices. However, the process comes with several challenges that can affect efficiency, collaboration, and overall development workflow. For example, long waiting times where PRs often remain unreviewed for extended periods, slowing down development. \citet{kudrjavets2022mining} reported that between 29\% and 63\% of the overall code review lifetime is spent waiting for the accepted changes to merge. \citet{bacchelli2013expectations} reveal that scheduling and time issues also appeared challenging; yet, understanding the code and comments takes most of the review time. According to these authors, code reviews tend to be less focused on detecting defects than expected and, instead, provide additional advantages, such as knowledge transfer, increased team awareness, and the creation of alternative solutions to problems. \citet{davila2021systematic} also reported that reviewers have found it difficult to understand the code, which prompts the need for enhancements to visualize code changes to facilitate understanding. However, a recent study has shown that ChatGPT is an alternative to generate some kind of explanations during MCR \cite{widyasari2023explaining}. \citet{davila2024tales} identified five challenges to adopting generative AI in MCR: trustworthiness, lack of context, misleading review comments, security and privacy, and token limits. \textit{Although prior studies have explored aspects of the use of GPT in software development, they do not provide a systematic analysis of its impact on waiting times and efficiency in all phases of the PR workflow. Our study addresses this gap by focusing on its influence during specific code review phases.}

\header{Factors Affecting PR Lifetime and Latency.}
Studying the factors that influence PR latency is crucial for improving software development efficiency, collaboration, and overall code quality. Understanding what causes delays in PR review and merging can lead to optimizations that benefit both developers and organizations \cite{zhang2022pull}, \cite{rose2017towards}, \cite{moreira2021factors}, \cite{khatoonabadi2024predicting}, and \cite{khatoonabadi2023wasted}. \citet{zhang2022pull} identified that the number of commits at open time is negatively correlated with PR merging but shifts to a positive correlation at close time. \citet{rose2017towards} built a Multiple Linear Regression Model (MLR) that indicates that the PR size has a statistically significant effect on the review time, as the positive value of the regression coefficient implies that as the PR size increases, the review time for PRs also increases. \citet{moreira2021factors} further emphasized that PRs possess some structural characteristics, such as the number of modified files, the number of commits, and the number of changed lines of code. \citet{moreira2021factors} concluded that the combination of structural characteristics may increase the chances of PRs having a very short and long lifetime.

Although prior research explores factors affecting PR latency (e.g., size, number of commits), there is limited research into how GPT can have an effect on PR resolution time. \citet{khatoonabadi2024predicting} developed a machine learning model to predict the latency of the initial responses of the maintainers and contributors after receiving the first feedback from the maintainer. These authors highlighted that the timing of submissions and feedback are factors that have a significant influence on response waiting times. Their approach, which provides estimated waiting times, helps open-source projects collaborate better by allowing maintainers and contributors to anticipate and address potential delays in PR reviews. \textit{This study complements the aforementioned work by offering empirical insights into GPT usage beyond commit messages, specifically examining detailed time reductions in key phases such as review and change.}

\header{The Application of LLMs in Code Reviews.}
LLM-powered tools can offer services designed to streamline various code review-related development tasks. Research has been conducted on the impact of AI-driven generative tools in this context \cite{tufano2024unveiling}, \cite{xiao2024generative}, \cite{davila2024tales}, \cite{rasheed2024ai}, \cite{vijayvergiya2024ai}, \cite{yu2024fine},  \cite{hao2024empirical}, \cite{das2024investigating}. For instance, \citet{davila2024tales} conducted a literature review on the application of AI-based generative tools in a code review context and highlighted that AI-based tools can positively impact code review efficiency by reducing reviewer workload and review duration and increasing the likelihood of acceptance in GitHub projects. \citet{xiao2024generative} observed that PRs assisted by Copilot have a 1.57 times higher likelihood of being merged compared to those not assisted by Copilot for PR. \citet{tufano2024unveiling} reported how developers use ChatGPT to automatically generate a commit message and a PR/issue description. \citet{rasheed2024ai} highlighted the potential of LLM-powered AI agents to considerably improve conventional code review procedures. These agents provide a twofold advantage: they enhance code quality and improve developer learning.

\citet{vijayvergiya2024ai} shared the experience in developing, implementing, and evaluating AutoCommenter, a LLM-powered code review tool that enforces coding best practices and has a positive impact on the developer's workflow. \citet{yu2024fine} proposed Carllm (Comprehensibility of Automated Code Review using LLMs) that improves the comprehensibility of automated code review. \citet{hao2024empirical} studied the role of ChatGPT in collaborative coding by analyzing the shared conversation between developers and ChatGPT in GitHub PRs and issues, finding that the most frequent requests involve code generation, conceptual understanding, how-to guidance, issue resolution, and code review. Studies on GPT have examined particular tasks (e.g., code generation, commit message generation, or bug fixing), but lack a holistic view of its phase-specific impacts during the PR life cycle. 

\citet{cihan2025automated}  conduct an empirical study on the use of LLM-based automated code review in the industry. They analyze 4,335 PRs across three projects. Although approximately 74\% of automated comments were applied, the tool slightly improved code quality, increased bug detection, and promoted best practices, but also led to longer PR closure times and occasional irrelevant suggestions. The study shows that automated reviews can help developers, but they do not replace human judgment. Effectiveness depends on the context of the project and current workflows. \citet{ramesh2025automated} present a case study on an LLM-augmented code review tool used by Ericsson. The tool combines static program analysis with prompt engineering to deliver context-aware reviews in a live workflow. Their evaluation suggests that integrated LLM-based reviews can reduce manual effort and support developer workflows. However, effective prompting, tool integration, and risk mitigation remain key challenges. \textit{Previous research has explored the use of large language models for automated code review in software development, but human involvement is still essential to ensure contextual understanding. This study aims to address this limitation by investigating the impact of GPT on specific phases of the PR workflow, specifically the review and the waiting phase. It is important to note that this study does not propose a new approach for automatically generating code reviews using LLMs.}
\section{Study Setup}
\label{sec:setup}

\begin{figure}
\centering
\includegraphics[width=\linewidth]{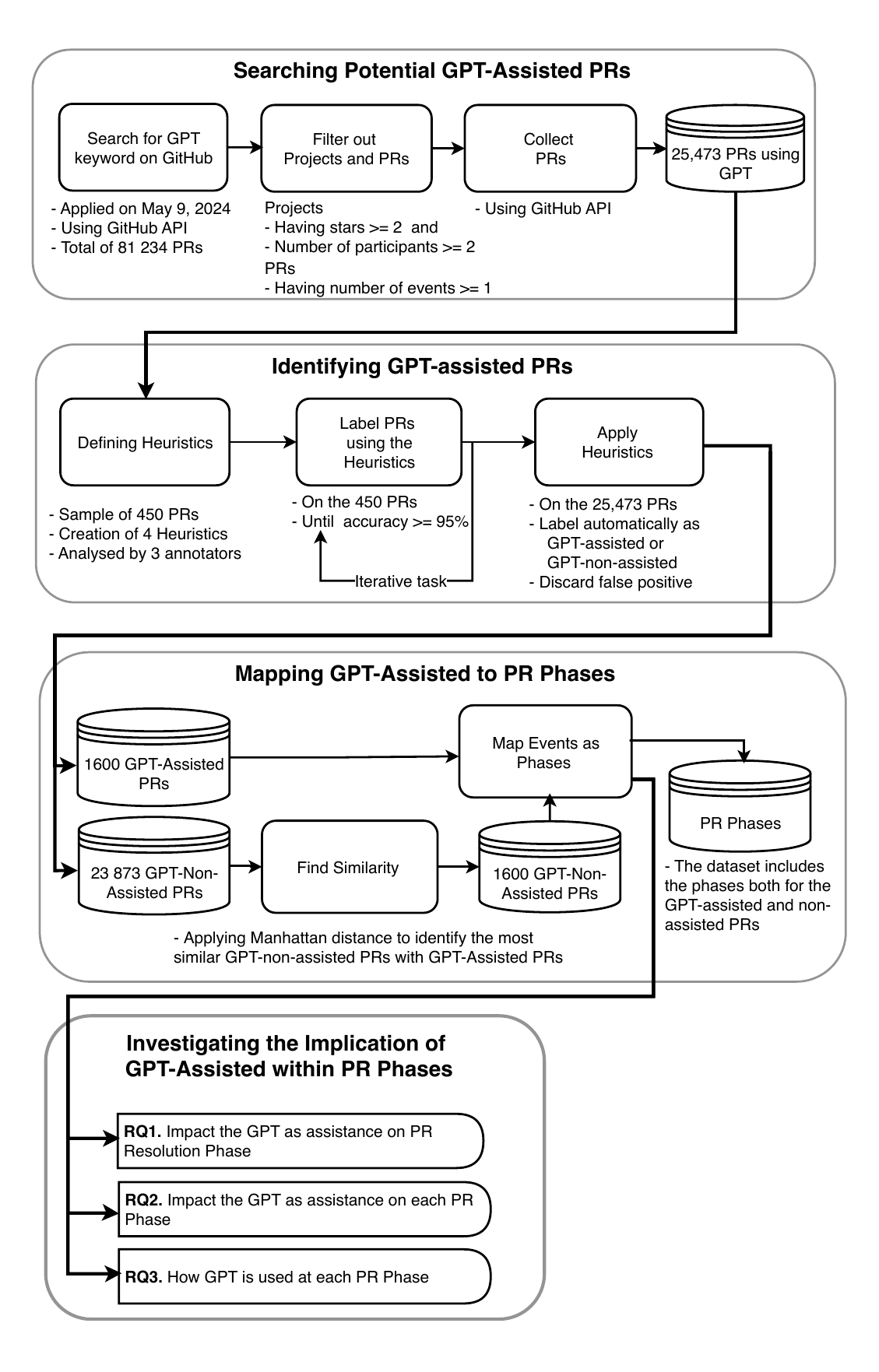}
\caption{An overview of our study setup that describes a semi-automated approach for identifying, detecting, and mapping GPT-assisted PRs to the phases in which GPT is used.}
\label{fig:study_setup_overview}
\end{figure}

In this paper, we investigate how incorporating GPT as an assistance tool influences the code review process. In particular, it focuses on scenarios where developers interact with GPT to obtain feedback or suggestions during PR reviews, while excluding PRs generated automatically by GPT. Specifically, we examine how GPT influences the overall PR resolution time and how its effects vary between different phases of the review workflow. We also aim to identify the types of review tasks for which developers actively seek GPT support. Figure~\ref{fig:study_setup_overview} presents an overview of our study design. We begin with an automated procedure to identify potential GPT-assisted PRs, followed by a semi-automated verification step to confirm the actual use of GPT during the review. Finally, we map the verified GPT-assisted PRs to their corresponding review phases. The remainder of this section provides a detailed explanation of each step in our methodology.

\subsection{Searching Potential GPT-assisted PRs}
\label{sec:setup-A}
To identify relevant PRs for our analysis, we rely on GitHub, which adopts a PR-based approach to submit, review, and merge code contributions~\cite{davila2021systematic}. We use the GitHub REST API~\cite{Github-API} to retrieve a broad set of candidate PRs in which the assistant of a GPT model could have been used during the review process.

\subsubsection{Searching} Following a similar strategy to that of ~\citet{tufano2024unveiling}, we conduct keyword-based searches for mentions of "GPT" in four main PR fields, obtaining a total of 81,234 PRs.

\begin{enumerate} 
\item[$\blacksquare$] \textbf{Project Name:} Although some GPT-focused projects may not actually use GPT for PR reviews, we still record these mentions before filtering out irrelevant cases \cite{pullrequest43}. 
\item[$\blacksquare$] \textbf{PR Title:} The use of GPT may be evident in the PR title (e.g., refactored using GPT suggestions \cite{pullrequest64}). 
\item[$\blacksquare$] \textbf{Body+Comments:} A PR description and/or comments can refer to GPT explicitly (e.g., code generated by GPT \cite{pullrequest16}).
\item[$\blacksquare$] \textbf{File(s) Name:} References in filenames (e.g., chatgpt\_review.md \cite{pullrequest1}) can indicate GPT-generated artifacts.
\end{enumerate}

\subsubsection{Filtering} 
\label{sec:setup-A-Filtering} 
To avoid projects with minimal activity or that lack meaningful collaboration, we keep only projects that meet the following criteria: (1)~ repositories that have at least 10 stars \cite{borges2018s}, to remove toys or trivial projects. (2)~at least two contributors are involved in the review process, to exclude self-reviewed PRs. (3)~PRs that contain more than one event in the lifetime.

Based on the aforementioned criteria, a total of 25,473 pull requests were collected from 9,254 repositories on May 9, 2024. For each PR, we extracted detailed metadata, including the associated project name, URL, PR number and title, event history, contributor information, creation date, comments, and the set of modified files.

\subsection{Identifying GPT-assisted PRs}
\label{sec:setup-B}
In this semi-automated step, our goal is to define the heuristics that indicate the use of GPT to classify PRs as GPT-assisted or GPT-non-assisted. By doing this, we exclude false positives (i.e., PRs that mentioned GPT as discussed in Section~\ref{sec:setup-A} but did not use GPT to assist in the code review process). This step ensures that the PRs used for our analysis are PRs where GPT models are used in the code review process, rather than just being part of the project code.

\subsubsection{Defining Heuristics} 
\label{sec:setup-B-Def-Heuristics}
To enable the identification of patterns related to GPT usage in code review processes, we randomly selected 450 GitHub PRs from the initial dataset with a confidence level of 95\% and a margin error of 5\%. We ensure an even distribution among the three contributing authors to examine indicators of GPT assistance during review activities.

Three authors independently inspected 150 unique PRs in three separate rounds, analyzing project names, PR titles, modified files, and PR bodies. In Round 1, each annotator reviewed the first set of 150 PRs and cross-validated our observations to ensure that the identified patterns consistently reflected GPT-assisted contributions. In particular, most PRs were ultimately classified as GPT-non-assisted, as GPT was often integrated into the software (e.g., through GPT APIs) but not explicitly used to support the code review itself. Table~\ref{tab:frequency_pattern_pr_r1} in the appendix presents the findings of Round 1.

In Rounds 2 and 3, the annotators applied the same procedure to two additional sets of 150 PRs each. After each round, the reviewers met to consolidate recurring patterns, clarify ambiguous cases, and resolve classification disagreements through discussion until consensus was reached. For relevance labeling, disagreements occurred in 4\% of the cases. Krippendorff's $\alpha = 0.77$ \cite{krippendorff2011} indicates substantial inter-annotator reliability. This suggests that the evaluators were largely consistent in their judgments, providing confidence in the quality of the relevance annotations.

We defined four heuristics, as outlined in Table \ref{tab:heuristics_definition}, to analyze and label the presence of GPT-assisted PRs. These heuristics are reinforced by a regular expression-based technique which automatically classifies the collected PR instances as either GPT-assisted or GPT-non-assisted based on the identified patterns: 
\textbf{H1:} Project name, 
\textbf{H2:} PR title, 
\textbf{H3:} Modified files, 
\textbf{H4:} PR body+comments.

\begin{table}
\caption{Definition of the Heuristics Applied to the Project Name, PR Title, PR Body, and Modified File to Identify PRs using GPT as Assistance}
\label{tab:heuristics_definition}
\begin{tabular}{p{0.6in}|p{3.8in}}\hline \hline
\textbf{Heuristic} &  \textbf{Description}  \\  \hline
H1 & Aims to find and filter any GPT-related keywords within the project name (e.g., EleutherAI/gpt-neo)  \\ \hline
H2 &  Matches and filters any pattern that implies the use of GPT as non-assistance in the PR title (e.g., use of chatgpt-retrieval-plugin) \\ \hline
H3 & Finds and filters any file name pattern, directory, URL, and file version that indicates the use of GPT for implementation (e.g., Add ChatGPT Image Unlocker V2, from 3.1.0 to 3.3.0) with the modified file names.  \\ \hline
H4 & Extracts and filters the patterns in H1, H2, H3 in the PR body, but including action verbs such as add, refactor, implement, and rename (e.g., Added some simple error handling around the OpenAI ChatGPT logic to capture rate limit errors and bad gateways).         \\ \hline
\end{tabular}
\end{table}

\subsubsection{Label PRs using Heuristics}
\label{sec:setup-B-Label_PR} Initially, we experimented with various sizes of n-gram to determine which configurations offered the most valuable insights for detecting recurring patterns. In contrast to the methodology employed by ~\citet{tufano2024unveiling}, which used 2- or 3-grams, we found that a maximum size of 5-n-gram provided sufficient context or relevant information to accurately capture the role of GPT in GPT-assisted pull requests. We then proceeded to automatically extract both forward and backward n-grams, up to a maximum of 5-grams \cite{jurafsky2023slp3}, containing the term \textit{GPT}, while simultaneously applying a set of heuristics designed to reduce the occurrence of false positives. For instance, when analyzing the PR body, we encountered the sentence: \textit{"Instead of incorporating muP into GPT-NeoX we are going to move these changes to [our fork of their repo]"}\footnote{https://github.com/EleutherAI/gpt-neox/pull/1061\#issuecomment-1771635901}. From this, we extracted the backward n-gram "of incorporating muP into GPT-NeoX" and the forward n-gram \textit{"GPT-NeoX we are going to"}; the application of heuristic \textbf{H4} identified \textit{GPT-NeoX} as a reference to the PR name, thereby classifying this PR as GPT non-assisted. Additionally, we applied heuristics on the sample of 450 PRs (defined in section~\ref{sec:setup-B-Def-Heuristics}) to filter and automatically eliminate false positives, including cases where GPT is mentioned in a PR but not used for assistance, as well as instances where a PR may appear GPT-assisted but is actually unrelated (e.g., mentions of GPT in the context of non-code-related discussions, such as research or tool integration). These measures ensured that only relevant PRs, where GPT is actively used for code review assistance, were included in the analysis.

To ensure the reliability of our heuristics, two authors independently examined the labeling on the 450 PRs by carefully reviewing the labels generated during the execution of the heuristics.  In the final step, the heuristics were adjusted and reapplied to improve accuracy and consistency. We performed three rounds of refinement/reapply until we obtained an accuracy of 95\% as an acceptable agreement score in the labeling of PRs. Table~\ref{tab:frequency_pattern_pr_percetage} in the appendix presents the increase in accuracy as more heuristics were refined. 

\subsubsection{Applying the Heuristics}
\label{sec:setup-B-Combining_Heuristics}
We applied the heuristics 
to the entire dataset of 25,473 PRs to label them as either GPT-assisted or non-assisted. Subsequently, one of the authors examined the dataset for true positive cases by manually reviewing the final labels. Ultimately, we split the dataset into two separate groups: \textit{GPT-assisted} and \textit{GPT-non-assistance} PRs. Of the 1,600 GPT-assisted PRs, 54 remain open, 179 have been closed without merging, and 1,367 have been successfully merged. This distribution highlights that the majority of GPT-assisted PRs are ultimately integrated into the codebase, reflecting the potential influence of GPT on the completion and acceptance of code contributions.
The table~\ref{tab:GPT-PR-Status-Summary} displays the distribution of GPT-assisted PRs in the dataset.

\subsection{Mapping Events to PR Phases}
\label{sec:setup-C}
In this automatic step, our goal is to define the different phases of the code review process as previously performed by \citet{kudrjavets2022mining}. We report descriptive statistics to quantify the extent of GPT usage in each phase, depending on whether non-productive phases are present in the review workflow.

\subsubsection{Finding Similarity GPT-non-assisted PRs with GPT-Assisted PRs} 
\label{sec:setup-C-similarity}
To have a fair comparison between the GPT-assisted and GPT-non-assisted PRs, we quantify the similarity between them  applying the Manhattan distance \cite{malkauthekar2013analysis, ye2021recommending} on the same factors as applied in RQ1 \ref{sec:results-rq1}. To mitigate the presence of outliers or bias in the factor, we rescale no\_commits, PR\_size, no\_changed\_files, project\_age to a range within [0, 1], applying the max-min method \cite{ye2021recommending}. The two well-known metrics (the Manhattan and Euclidean distance produce very similar results (Spearman correlation \(\rho=0.98\) \cite{JSSv094i12} when computing the distances between factors), we finally test the differences between the PRs supported by or without GPT by getting the minimum distance value from the Manhattan distance.

\subsubsection{Identifying PR Events using GPT} 
\begin{figure}
\centering
\includegraphics[width=\linewidth]{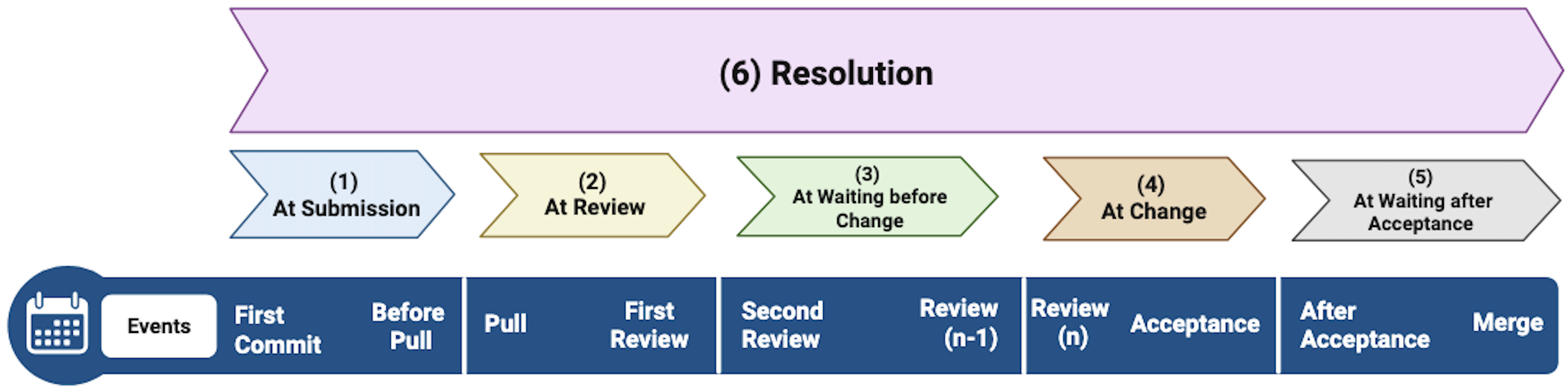}
\caption{Pull Request Phases}
\label{fig:Pull-Request-Phases}
\end{figure}

For each PR previously labeled as GPT-assisted, we applied the heuristic H4 (explained in Table \ref{tab:heuristics_definition}) to timeline events \cite{GitHubTimeline} to label the events that the contributor and reviewers shared as GPT-assisted or GPT-non-assisted events. This specifies when and in what capacity GPT helps with PR discussions. Figure \ref{fig:Pull-Request-Phases} is a visual representation of our definition of the PR phases:

\begin{enumerate}
\item[$\blacksquare$] \textbf{at\_submission:} the time in hours from the first commit (committed state) to when the author submits the code change (pulled state). According to Figure \ref{fig:Pull-Request-Phases}, this phase corresponds to the time taken from \(at\_commit_{event=0}\) to \(at\_before\_pull\).

\item [$\blacksquare$] \textbf{at\_review:} is the amount of time, expressed in hours, that occurs between the submission of a code review (pulled state) and the first review, comment, inline comment (comment on specific code fragments), commit-commented, review-commented, provided by a person who is not the code's author. The review phase is comparable to the time-to-first-response, which is a proportion of the time needed to accomplish the first iteration of the code review. According to Figure \ref{fig:Pull-Request-Phases}, this phase comprises the period between from \(at\_pull_{event!=0}\) to \(at\_review_{n=1}\).

\item [$\blacksquare$] \textbf{at\_change:} refers to the time in hour from the last review, comment, inline comment (comment on specific code fragments), commit-commented, review-commented until the acceptance of the code review by someone other than the author of the code. According to Figure \ref{fig:Pull-Request-Phases}, this phase covers the time between \(at\_review_{n}\) and \(at\_accepted\).

\item[$\blacksquare$] \textbf{at\_waiting\_time:} 
refers to the duration, in hours, spent in iterative exchanges and feedback between reviewers and the contributor. The first interval (waiting before change) captures the time between  \(at\_review_{2}\) and \(at\_review_{n-1}\), while the second (waiting after acceptance) represents the period from acceptance to merge. During the waiting phases, several iterative exchanges occur between the contributor and the reviewer.
\end{enumerate}

It is important to note that when a second review \textit{(event = reviewed)} is absent in the PR timeline, the PR transitions directly to the \textit{At Change phase}, indicating that no multiple back-and-forth interactions occurred between the contributor and the reviewers. Furthermore, We did not identify any events that indicate any GPT-assisted activity during the \textit{Submission and  Waiting after Acceptance} phases; therefore, these phases were excluded from further analysis in our study.

To analyze the impact of LLMs on each PR phase, we mapped a chronological sequence of events (including pulling, examining, integrating, forcing the head reference, commenting on reviews, comments in lines, commit messages, and merging) to the phases. To maintain the validity of our results, only pull requests with sequential timestamps were included in the analysis. In these cases, the review phase follows the submission phase, the approval step follows the review, and the merging occurs after acceptance. Finally, to ensure human participation in the review process, we excluded PRs in which the only participants, either as contributors or reviewers, were the original author and automated bots.

\section{Results}
\label{sec:results}

\subsection{\textbf{\rqi}}
\label{sec:results-rq1}
\header{Motivation:} 
Recent studies have investigated the influence and popularity of ChatGPT on code review \cite{guo2024exploring}, \cite{grewal2024analyzing}, \cite{hao2024empirical}, \cite{watanabe2024use}. The potential impact of using GPT models as assistants throughout the entire PR lifecycle remains largely unexplored. This research question investigates whether relying on GPT-assisted PRs affects code review resolution time in GitHub projects by analyzing the impact of GPT models on the overall merge time of PRs.

\header{Approach:} To answer this research question, we build a Multiple Linear Regression Model (MLRM) \cite{james2013introduction} to empirically explain PR lifetime. For this purpose, we examine previous studies to identify a list of known factors that significantly influence PR resolution time \cite{zhang2022pull}, \cite{moreira2021factors}, \cite{rose2017towards}, \cite{silva2020measuring}. \citet{bernardo2018studying} defines resolution time as the time it takes for a code change to be merged into the development branch after review. This implies that the change is not fully implemented until it can be built, tested, and deployed. This is possible only after merging into the main branch. We will use the time it takes for code modifications to be merged as a metric, selecting only PRs that have been merged, considering the identified factors that affect this time \cite{silva2020measuring}. After this step, our dataset includes only the merged PRs with a total of 1,367 PRs (85.43\% of the total GPT-assisted PR population as shown in Table~\ref{tab:GPT-PR-Status-Summary}) using GPT-assisted. To compare the resolution time between the GPT-assisted PRs and GPT-non-assisted PRs, we used the GPT-non-assisted PRs dataset explained in section~\ref{sec:setup-C-similarity} (Finding Similarity GPT-non-assisted PRs with GPT-Assisted PRs). Furthermore, 
we added a two-state indicator to the dataset, with 1 representing PR that use GPT-assisted and 0 representing PR that use GPT non-assisted. This indicator is used to better understand the impact of the GPT as a factor on PR latency.

To compare the statistical difference between PRs using GPT-assisted or GPT-non-assisted, as an influencing factor, we use the Mann-Whitney U test, a non-parametric test that compares two independent groups that do not require normally distributed samples\cite{nachar2008mann}. To carry out this statistical test, we utilize the stats package\footnote{R Core Team. 2023. R: a language and environment for statistical computing. https://www.R-project.org} considering a confidence level of 95\% (\(\alpha=0.05\)). The following formula and list of factors characterize the Linear Model (LM) performed in our study:

\begin{equation}
\begin{aligned}
time\_to\_merge & = is\_gpt\_assisted  \\
& + no\_commits \\ 
& + PR\_size \\
& + no\_changed\_files \\
& + project\_age 
\end{aligned}  
\label{Equation 1}
\end{equation} 

\begin{enumerate}
\item [$\blacksquare$]
\textit{time\_to\_merge (dependent variable)} is a numeric value that indicates the latency of PR in hours.
\item[$\blacksquare$] \textit{is\_gpt\_assisted (Treatment)} is a binary value ([0,1]) that translates whether a PR is a GPT-assisted PR. 
\item[$\blacksquare$] Common factors that influence the PR latency \cite{zhang2022pull,rose2017towards,moreira2021factors}.
\begin{enumerate}
\item[$\square$] \textit{no\_commits (control variable) } represents the number of changes pushed to a repository over time.
\item[$\square$] \textit{PR\_size (control variable)} is the amount of change in the PR (added \(+\) deleted lines of code).
\item[$\square$] \textit{no\_changed\_files (control variable)}  is the number of files that have changed during the review process.
\end{enumerate}
\item[$\blacksquare$] The \textit{project\_age in months (control variable)} variable, which represents the maturity level of the projects.
\end{enumerate}

To mitigate the influence of skewness and enhance the clarity of our statistical examination, we implemented a logarithmic transformation on variables that exhibit a skewed distribution \cite{feng2014log}. To implement our linear model shown in Equation \ref{Equation 1}, we follow the approach proposed by \citet{james2013introduction}. First, we build our model using the lm function that comes standard with the distribution\footnote{https://www.rdocumentation.org/packages/stats/versions/3.6.2/topics/lm}. Then we evaluate the degree of confidence of our model by obtaining the marginal and conditional coefficients of determination \((R^2)\) \cite{nakagawa2017coefficient}. To determine whether the adoption of GPT-assisted has a potential impact on PR latency, we check the statistical significance and the total variance of \textit{is\_gpt-assisted} in the model. To identify the level of significance of the variables, we rely on the traditional confidence level of \(\alpha=0.05\) \cite{chavalarias2016evolution}. 

Following the methodology of ~\citet{cole2017statistics}, we quantified the \textit{percentage reduction} between GPT-assisted and non-assisted PRs over the entire PR lifecycle (i.e., resolution or merge time), expressing the difference relative to the median merge time of non-assisted PRs:

\begin{equation}
\begin{aligned}
(\text{1} - \frac{\text{Median GPT-assisted PRs merge time}}{\text{Median GPT-non-assisted PRs merge time}}) \times 100 
\end{aligned}  
\label{Equation 2}
\end{equation} 

\begin{figure}[!h]
\centering
\includegraphics[width=\linewidth]{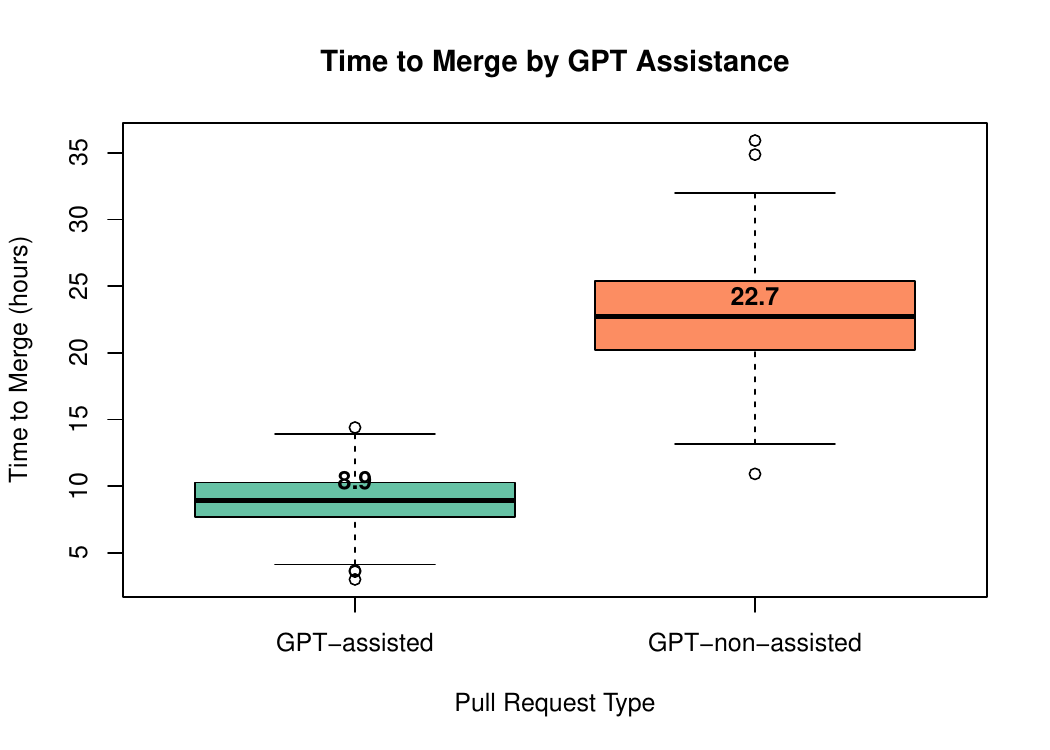}
\caption{The distribution of merge times reveals that GPT-assisted PRs have a median merge time of approximately 9 hours, compared to 23 hours for non-assisted PRs.}
\label{fig:Mean time to merge at the Resolution phase}
\end{figure}

\header{Findings:} As shown in Table \ref{tab:lm_coefficients}, the multiple linear regression model suggests that GPT assistance has a statistically significant effect on pull request (PR) latency. Specifically, the variable \textit{is\_gpt-assisted} yields a \textit{p-value of $4.88e^{-11}$}, well below the standard significance threshold of \(\alpha=0.05\), indicating strong evidence against the null hypothesis. The corresponding negative coefficient suggests that GPT assistance is associated with a reduction in the time required to merge a PR. The standard error for this estimate is 0.0745, indicating a high degree of precision, and the large associated t-value reinforces the robustness of the finding. Together, these results demonstrate that GPT-assisted PRs tend to be merged more quickly than their non-assisted counterparts. This quantitative result is consistent with the descriptive statistics presented in Figure \ref{fig:Mean time to merge at the Resolution phase}, where GPT-assisted PRs have a median merge time of approximately 9 hours, compared to 23 hours for non-assisted PRs.

\begin{table}
\centering
\caption{Linear Model Coefficients for log1p(time\_to\_merge\_h)}
\label{tab:lm_coefficients}
\begin{adjustbox}{width=\columnwidth,center}
\begin{tabular}{lrrrrl}
\hline
\textbf{Predictor} & \textbf{Estimate} & 
\textbf{Std.Err} & \textbf{t value} & 
\textbf{Pr($>|$ t $|$)} & \textbf{Signif.} \\
\hline
(Intercept) & 0.5186   & 0.1248  & 4.154  & 3.37e-05  & *** \\
is\_gpt-assisted                 & -0.4921  & 0.0745  & -6.601 & 4.88e-11  & *** \\
log1p(no\_commits)        & 0.8052   & 0.0715  & 11.267 & $<$ 2e-16 & *** \\
log1p(PR\_size)           & 0.1576   & 0.0322  & 4.899  & 1.02e-06  & *** \\
log1p(no\_changed\_files) & 0.0064   & 0.0702  & 0.091  & 0.927     &     \\
log1p(project\_age\_month) & 0.3371   & 0.0275  & 12.252 & $<$ 2e-16 & *** \\
\hline
\multicolumn{6}{l}{Signif. codes: 0 ‘***’ 0.001 ‘**’ 0.01 ‘*’ 0.05 ‘.’ 0.1 ‘ ’ 1} \\
\end{tabular}
\end{adjustbox}
\end{table}

\begin{tcolorbox}[ standard jigsaw, opacityback=0]
\textbf{Answer to RQ1.} 
The results highlight the potential of GPT assistance to significantly accelerate the pull request review process, with percentage of reduction of the merge times by 61\%. Integrating GPT as an assistance in the development workflows can help streamline code review, improve team efficiency, and facilitate faster software delivery.
\end{tcolorbox}

\subsection{\textbf{\rqii}}
\label{sec:results-rq2}
\header{Motivation:}
Previous studies have shown that factors such as technology, collaboration dynamics, and project context influence the effectiveness of code reviews \cite{zhang2022pull, kudrjavets2022mining, baysal2016investigating, bernardo2018studying}. More recently, AI-driven tools, particularly LLMs, have been explored for their potential to assist with PR-related development tasks and improve review efficiency \cite{davila2024tales, xiao2024generative, tufano2024unveiling}. Despite growing interest, little is known about how developers actually use GPT during the code review process or how its usage varies across different phases of a PR lifecycle. This gap motivates our investigation of the impact of GPT as assistance on review and resolution time. By identifying the specific phases in which GPT is applied and understanding its role as an assistive (rather than an automated) tool, we aim to assess where it offers the greatest benefit in the review workflow.

\header{Approach:} To investigate the impact of using GPT as assistance throughout the PR phases, we first define the distinct phases of the PR workflow, starting from the submission of a change to merge into the project branch including the presence of the waiting time, as described in detail in the section ~\ref{sec:setup-C} (Mapping Events to PR Phases). To assess whether the use of GPT-assisted leads to statistically significant differences in PR phases, the Mann-Whitney U test was applied \cite{mann1947test}. This non-parametric test is well-suited for comparing two independent samples, particularly when the assumption of normality cannot be guaranteed. In this study, the GPT-assisted and GPT-non-assisted datasets are treated as independent groups, as each PR belongs to only one condition (each PR is placed into only one of these two groups), with no overlapping or paired observations. Using the Mann-Whitney U test aims to determine whether the distributions of phases differ significantly between the two groups, thus providing evidence on the potential impact of GPT-assisted. We computed percentage reductions in PR phases between GPT-assisted and GPT non-assisted cases, following the definition of ~\citet{cole2017statistics} as defined in Formula~\ref{Equation 2}. To account for the asymmetry of the raw percentage measures, we also calculated percentage of reduction in logarithmic transformation, confirming a substantial time benefit associated with GPT assistance:
$-109.9\%$ and $-208.7\%$ reductions for the \textit{At Review} and \textit{At Waiting for Change} phases, respectively.

\header{Findings:} The results presented in Table~\ref{table:RQ2-median} highlight significant differences in the duration of review between GPT-assisted and non-assisted PRs, with particularly notable reductions observed during the \textit{At Review} and \textit{At Waiting Before Change} phases. In the \textit{At Review} phase, the median time for GPT-assisted PRs was approximately 1 hour, compared to 3 hours for GPT-non-assisted PRs — reflecting a 66.7\% reduction in the review time. The most significant difference was found in the \textit{At Waiting Before Change} phase, where GPT-assisted PRs had a median waiting time of approximately 3 hours, while their non-assisted counterparts experienced a median of 24 hours. This corresponds to an 87.5\% reduction in waiting time. The results of the Mann–Whitney U test confirmed that the reductions in these two phases were statistically significant, with p-values well below the conventional significance threshold (\(\alpha\) = 0.05). These low p-values provide strong evidence against the null hypothesis, suggesting that the differences are unlikely to be due to random variation.
Interestingly, no significant differences were observed in Phase 1 (\textit{At Submission}) and Phase 5 (\textit{At Waiting after Acceptance}), where there was no clear indication of GPT as assistance was used during these stages, suggesting that the use of GPT is concentrated in the core iterative phases of the review process.

\begin{table}
\centering
\caption{Median time (in hours) spent across three key PR phases for GPT-assisted and non-assisted pull requests}
\label{table:RQ2-median}
\begin{adjustbox}{width=\columnwidth,center}
\begin{tabular}{cccc}\hline
\hline
\textbf{GPT} 
& \textbf{At Review}
& \textbf{At Waiting for Change}
&  \textbf{At Change}  \\
Assisted &  1 & \cellcolor{green!25} \textcolor{orange}{3} & 1  \\
Non-Assisted & 3  & \cellcolor{red!25}\textcolor{blue}{24} & 1 \\
\hline 
\end{tabular}
\end{adjustbox}
\begin{flushleft}
\footnotesize{Note: There is no evidence of GPT assistance being used during the submission or waiting after acceptance phases of the PR workflow. Consequently, these phases are excluded from the table.}
\end{flushleft}
\end{table}

\begin{tcolorbox}[ standard jigsaw, opacityback=0]
\textbf{Answer to RQ2.} Our analysis reveals that GPT-assisted PRs are associated with significantly shorter durations in specific phases. The review phase is 66.7\% faster, and the revision before acceptance is reduced by 87.5\% compared to GPT-non-assisted PRs. However, no significant differences are observed in the change phase.
\end{tcolorbox}

\subsection{\textbf{\rqiii}}
\label{sec:results-rq3}
\header{Motivation:}
Our prior analysis revealed that the use of GPT was associated with notable variations in the duration of individual PR phases, suggesting that its influence extends beyond the overall resolution time to phase-specific developer activities. While earlier studies have examined how developers use GPT for specific code review tasks \cite{tufano2024unveiling, chouchen2024so}, recent work has raised concerns about its influence on review discussions and developer responses \cite{hao2024empirical, watanabe2024use}. Although GPT tools show promise across various software engineering activities, little is known about their role across different phases of the PR workflow. This suggests an underexplored but potentially broad impact of GPT-based tools on the code review process. To fill the gap, we analyze the purposes for which developers use GPT within GitHub PR phases. 

\header{Approach:}
To examine the purposes for which developers use GPT as assistance, we analyzed a statistical sample of 310 randomly selected PRs with a confidence level of 95\% and a margin error of 5\%, drawn from the dataset of 1,600 GPT-assisted PRs as described in Section~\textit{\ref{sec:setup-B-Combining_Heuristics}}. The labeling followed the taxonomy proposed by~\citet{tufano2024unveiling} and was conducted in two rounds of 155 PRs each. Two authors independently assigned labels (\textit{enhancement, implementation, bug fix, testing, documentation, other}) to the instances, then met after each round to refine definitions. 
The authors resolved any discrepancies or disagreements in the labeling through discussion and consensus, rather than the formal calculation of inter-rater reliability metrics
This  approach allowed the authors to assign labels consistently and accurately even in nuanced cases. The annotators agreed that in some cases, multiple labels should be applied to one task, even if it is considered "atomic". The analysis specifically targeted events in which developers explicitly recorded their requests to GPT and the corresponding suggestions generated by the model. This classification enabled a structured understanding of the functional contexts in which GPT-assisted PRs are integrated into the PR workflow.

\header{Findings:}
Table \ref{tab:RQ3-tasks} presents the most common tasks for which developers seek the assistance of GPT during each phase of the PR process, along with details on their frequency and significance, highlighting its focused role in supporting diverse development tasks. In the following, we provide a detailed description of these tasks and their distribution across the different PR phases, highlighting the overall use of GPT on the code review lifecycle.

\begin{enumerate}
\item [$\blacksquare$] \textbf{Enhancement:} 
As shown in Table \ref{tab:RQ3-tasks}, this is the most common activity during the stages (60.26\% At Review, 42.31\% Waiting before Change and 35.90\% At Change), when developers offer current code as a baseline for adaptation. Most of the suggestions made by GPT are about refactoring the code, missing error handling, and making changes to existing code for performance reasons. We can see that the rewriting of the code was for minor modifications; for example, the most common modification was renaming a variable, function or class name, changing a flag variable, declaring a variable, adding/removing a conditional or reorganization a function. In contrast, we see more occasions when GPT proposes adding an error-handling instruction (try..catch). An example of a case of enhancement can be found in the PR \cite{pullrequest2059}, where the developer asked ChatGPT to change a Ruby plugin:
\textit{updates external-posts.rb plugin, allowing the user to specify an explicit lists of urls in config.yml that are then displayed in the blog feed as external posts.
99\% of the code in this change is written by gpt-4:} 
\url{ https://chat.openai.com/share/24432d24-36a7-4d6f-a5c0-d7e5142f68cd}

\item [$\blacksquare$] \textbf{Implementation:} Contrary to our expectations, in our sample we observe that the \textit{Implementation} in which the developers ask GPT to generate a piece of code (from scratch) has a low frequency, as shown in Table \ref{tab:RQ3-tasks} where the ratio is split on 6.41\% At Review, 10.26\% Waiting before Change and 8.97\% At Change. An illustrative instance is presented in \cite{pullrequest1858}, where a developer sought assistance from GPT in generating code to retrieve remote resources, and GPT responded with a Python-based implementation. 
We identified several cases in which reviewers expressed disagreement with the solutions generated by GPT. One such example is documented in \cite{pullrequest188}, where the reviewer remarked, \textit{"I will not blindly accept any AI-generated code"} and \textit{"using GPT so might not work as intended."}. 

\item [$\blacksquare$] \textbf{Bug fixing:}. This task ranks as the second most common use case for GPT assistance, occurring primarily during the Review phase (25.64\%) and to a lesser extent during the Change phase (21.80\%). In many instances, developers consult GPT to diagnose why a specific piece of code is not functioning correctly or to request help in fixing an error (e.g., \textit{“GPT found 2 config files not exist in the dataset list.”}) \cite{pullrequest1171}. GPT-generated suggestions also include improvements such as adding error-handling logic —for example: \textit{“The code does not seem to have any error handling. This could lead to unexpected behavior if an error occurs. Consider adding try-catch blocks to handle potential errors.”} \cite{pullrequest238}.

\item [$\blacksquare$] \textbf{Testing:} We find few cases where developers ask for GPT-assisted about \textit{testing} related tasks in our sample (\textit{every single character of tests is generated by GPT-4}) \cite{pullrequest4616}, (\textit{I have added the shorthand option -f for the --files flag in the \_ask.py file and updated the test case in test\_gpt\_cli.py.})\cite{pullrequest43}. 
We also notice that the \textit{testing} inquiries to GPT were for more complex tasks such as classes and core logic testing  (\textit{Chat GPT says this might work: the class AZDemoContentTest extends QuickstartFunctionalTestBase...})\cite{pullrequest3108}. 

\item [$\blacksquare$] \textbf{Documentation:} We observe that developers utilize GPT both for generating new documentation — such as implementation descriptions in markdown files or tutorials on how to use specific functionalities \cite{pullrequest109} — and for modifying existing documentation, including improving navigability, enhancing readability and formatting, and updating outdated links \cite{pullrequest28300}. Notably, in most instances, GPT-generated suggestions pertain to the README (.md) file \cite{pullrequest36, pullrequest736}.

\item [$\blacksquare$] \textbf{Other tasks:} We notice that \textit{At Changed Phase and At Waiting before Change}, non-programming related tasks count for the most requested activity with 28.20\% and 17.95\% respectively.  Among the most common questions developers ask GPT for suggestions are about translation, file versions, and setting up CD/CI workflows (YAML files, file configuration, Docker, Kubernetes), Copyright, and User Interfaces \cite{pullrequest27905,pullrequest1122,pullrequest31}. 
\end{enumerate}

\begin{table}
\centering
\caption{Frequency of Tasks That Developers Ask GPT for Assistance in PR Phases Within a Sample of 310 GPT PRs}
\label{tab:RQ3-tasks}
\begin{adjustbox}{width=\columnwidth,center}
\begin{tabular}{rrrr}\hline
\hline
\textbf{Tasks} 
& {\centering \textbf{At Review \% }}   
& {\centering\textbf{At Waiting for change \%}} 
& {\centering \textbf{At Change \%}}  \\

Enhancement &  \cellcolor{red!25}\textcolor{blue}{60.26} &\cellcolor{yellow!25} \textcolor{orange}{42.31} & \cellcolor{green!25}\textcolor{red}{35.90}  \\
Implementation & 6.41 & 10.26 & 8.97  \\
Bug Fix & \cellcolor{red!25}\textcolor{blue}{25.64} & 15.38 & \cellcolor{green!25}\textcolor{red}{21.80}  \\
Testing & 1.28 & 0.00 & 5.13  \\
Documentation & 11.54 & 12.82 & 10.26  \\
Other & 12.82 & \cellcolor{yellow!25} \textcolor{orange}{17.95} & \cellcolor{green!25}\textcolor{red}{28.20}  \\
\hline 
\end{tabular}
\end{adjustbox}
\begin{flushleft}

\footnotesize{
Note: Tasks may overlap across phases. A single GPT-assisted PR can involve multiple task types (e.g., enhancement and bug fix) and appear in more than one phase (review, waiting for change, or change). Therefore, percentages are not mutually exclusive and may sum to more than 100\%.}
\end{flushleft}
\end{table}

\begin{tcolorbox}[ standard jigsaw, opacityback=0]
\textbf{Answer to RQ3.} Contributors and reviewers mainly use GPT to assist with coding enhancements, bug fixes, and documentation. It is particularly useful during the review phase, helping to identify and resolve issues, refine code, and improve documentation. However, its use for creating new features or writing tests is less common. This pattern indicates that GPT primarily functions as a support tool for enhancing existing code rather than creating new components. This contributes to the effectiveness and quality of the code review process.
\end{tcolorbox}

\section{Discussion}
\label{sec:discussion}

\subsection{Implications for Practitioners}
\label{sec:discussion-1}
These findings have important implications for practitioners. Developers can integrate GPT-assisted PR into their phases to accelerate code reviews, reduce latency, and improve collaboration. In doing so, organizations can streamline their code review processes and focus on delivering value more efficiently. In addition, establishing best practices for the use of GPT-assisted can further enhance its effectiveness and ensure seamless adoption across teams. By integrating GPT into the pull request (PR) process, teams can significantly reduce review times, expedite the implementation of changes, and minimize waiting periods, ultimately streamlining their code review cycles. This efficiency gain allows developers to focus more on delivering value rather than being hindered by prolonged code review timelines. To maximize the effectiveness of GPT, teams should consider developing best practices for its use. For example, defining specific scenarios where GPT as assistance is most beneficial, such as refactoring code, resolving bugs, and enhancing documentation, can help ensure its targeted and efficient application. Additionally, encouraging collaboration between developers and reviewers when leveraging GPT can help maintain high-quality outcomes and prevent over-reliance on automated tools.

\subsection{Implications for Researchers}
\label{sec:discussion-2}
An important avenue for future research lies in examining the risks of user over-reliance on GPT in code review processes. While generative AI tools like ChatGPT offer remarkable potential to streamline code review phases and enhance productivity, there is a need to understand how unproductive interactions with ChatGPT might emerge and how these could undermine efficiency or decision-making. Identifying patterns of ineffective GPT use can help researchers propose strategies to optimize its integration into code review practices in real-world scenarios. Future studies could focus on designing frameworks and interfaces that guide users toward more effective use of GPT while maintaining control of their code review workflow. For instance, developing tools that provide metrics on the quality (i.e., reliability, usability, maintainability, testability of the code generated), confidence, or relevance of GPT’s solutions, enabling users to make more informed decisions and ensuring that developers maintain direction and avoid veering into unproductive tasks. 

Moreover, researchers could explore the development of features that encourage reflective practices, such as visualizations that compare GPT’s suggestions with best practices or community standards. These features could help users critically evaluate AI-generated solutions rather than passively adopting them. Such studies should examine their impact on developer productivity, decision-making, and team dynamics over time. Longitudinal studies could provide valuable insights into how extended use of GPT affects user skill retention, adaptability, and overall collaboration in software engineering.

\section{Threats to Validity}
\label{sec:threats}
In this section, we discuss the threats to the internal and external validity of our study.

\subsection{Construct Validity}
\label{sec:tv-1}
Construct validity concerns whether the study measurements accurately capture the concept of GPT assistance and its effects on the PR review process. In this study, GPT use was identified through explicit mentions of the keyword “GPT” in PRs, which may introduce false negatives (when GPT assistance is used but not explicitly mentioned) and false positives (when “GPT” appears in unrelated contexts). This limitation could bias the classification of PRs as GPT-assisted and affect the precision of the findings. Although a semi-automated verification step mitigated this issue, future work could employ more robust NLP-based intent analysis for detection. Another threat arises from how GPT assistance is operationally defined. The study uses heuristic rules to label PRs as GPT-assisted or non-assisted, which may lead to misclassifications when contributors do not document GPT use explicitly. To address this, manual inspections were conducted to refine the heuristics, achieving high labeling precision (95\%), and substantial inter-rater agreement was confirmed using Cohen’s Kappa, with discrepancies resolved by consensus among authors. The measurement of PR phases (e.g., review and waiting) using GitHub timestamps and events provides an objective foundation but may overlook qualitative aspects such as interaction complexity or the nature of the contributions by GPT. Inconsistencies in event logging across projects were mitigated by cross-checking phase definitions and excluding anomalous data. Furthermore, attributing observed reductions in PR phase durations solely to GPT assistance remains challenging, as other factors, such as task simplicity or team dynamics, could also influence outcomes. Although linear modeling accounted for some confounders (e.g., project age, number of commits), residual confounding may persist. Finally, developer interactions with GPT are not fully captured by temporal metrics; aspects such as behavioral or cognitive effects require further qualitative exploration. Future studies could integrate developer surveys or interviews to deepen understanding of how GPT shapes collaboration and review practices.

\subsection{Internal Validity}
\label{sec:tv-2}
Internal validity addresses the extent to which the observed effects can be causally attributed to GPT assistance rather than other influencing factors. Several potential confounding variables, such as code change complexity, developer experience, and team workflow dynamics, can have affected PR resolution time independently of GPT use. Although the study controls for some of these confounders using regression models, residual effects could still persist. Future research could strengthen control through methods such as matching or propensity score analysis to account for unobserved variables. Another concern relates to task categorization since developers may interpret or record their use of GPT differently, leading to inconsistencies in how tasks are classified. Such subjectivity could skew results and obscure patterns regarding the types of assistance most frequently used. Finally, while the study observes reduced time across various code review phases for GPT-assisted PRs, it remains crucial to verify that these reductions are genuinely attributable to the contributions of GPT rather than from external factors such as more efficient developers or inherently simpler tasks. Ensuring that time savings reflect true GPT influence is essential to maintaining the internal validity of the findings.

\subsection{External Validity}
\label{sec:tv-3}
External validity concerns the extent to which the findings of this study can be generalized beyond the specific context analyzed. Since the dataset is derived exclusively from GitHub, the results may be influenced by platform-specific characteristics, such as community norms or review workflows, that differ from those on other platforms like GitLab or Bitbucket. The study also focuses on repositories with at least ten stars, ensuring a baseline level of activity but potentially introducing selection bias, as less popular or private repositories may exhibit different development and review practices. Additionally, the wide variability among PRs (in terms of code size, complexity, and task type) poses challenges to generalization, as these factors can affect both the use and impact of GPT assistance. To enhance external validity, future research could incorporate data from multiple platforms, include a broader range of repositories, and apply stratified sampling or clustering techniques to better capture task diversity and contextual differences
\section{Conclusion}
\label{sec:conclusion}
This study highlights the potential of generative AI tools, particularly LLMs such as ChatGPT, to enhance collaborative software development. Through an analysis of their application in GitHub PRs, we observe substantial improvements across key phases of the PR workflow, most notably within the review phase, which exhibits a 66.7\% reduction in review time and an 87.5\% decrease in waiting time for acceptance. These results show the capacity of GPT to improve code review efficiency and support diverse tasks, including code enhancement, bug resolution, and documentation generation, with the most significant gains occurring during the review process. While the contributions of GPT are clearly beneficial in certain phases, its limited influence during the submission and post-acceptance phases points to areas for further development and optimization. Future research should investigate how GPT can be more effectively integrated across the entire PR lifecycle and evaluate its impact on long-term productivity. As LLMs continue to advance, they are poised to become essential components of software engineering, promoting greater efficiency, collaboration, and code quality. By harnessing the capabilities of these tools, both practitioners and researchers can shape the evolution of collaborative development workflows and drive innovation in the field of software engineering.

\section*{Author Contributions}

\textbf{Antonio Collante:} Conceptualization, Methodology, Software, Investigation, Data Curation, Writing Original Draft, Writing - Review and Editing, Visualization
\textbf{Samuel Abedu:} Conceptualization, Software, Data Curation, Writing - Review and Editing
\textbf{SayedHassan Khatoonabadi:} Conceptualization, Methodology, Validation, Writing - Review and Editing, Project Administration
\textbf{Ahmad Abdellatif:} Conceptualization, Validation, Writing - Review and Editing
\textbf{Ebube Alor:} Data Curation
\textbf{Emad Shihab:} Conceptualization, Supervision, Project Administration, Funding Acquisition

\section*{Acknowledgements}
This work was supported by the NSERC CREATE grant number 555406, 2021.

\bibliographystyle{elsarticle-num-names}
\bibliography{bibliography}

@misc{Github-API,
  title={GitHub REST API documentation},
  URL={https://docs.github.com/en/rest?apiVersion=2022-11-28},
  year={2024},
  consulted={2024-02-01}
}

@inproceedings{tufano2024unveiling,
  title={Unveiling ChatGPT’s Usage in Open Source Projects: A Mining-based Study},
  author={Tufano, Rosalia and Mastropaolo, Antonio and Pepe, Federica and Dabi{\'c}, Ozren and Di Penta, Massimiliano and Bavota, Gabriele},
  booktitle={2024 IEEE/ACM 21st International Conference on Mining Software Repositories (MSR)},
  pages={571--583},
  year={2024},
  organization={IEEE}
}

@article{zhang2022pull,
  title={Pull request decisions explained: An empirical overview},
  author={Zhang, Xunhui and Yu, Yue and Gousios, Georgios and Rastogi, Ayushi},
  journal={IEEE Transactions on Software Engineering},
  volume={49},
  number={2},
  pages={849--871},
  year={2022},
  publisher={IEEE}
}

@phdthesis{rose2017towards,
  title={Towards Understanding What Factors Affect Pull Request Merges},
  author={Rose, Tresa},
  year={2017},
  school={Carleton University}
}

@article{moreira2021factors,
  title={What factors influence the lifetime of pull requests?},
  author={Moreira Soares, Daricelio and de Lima J{\'u}nior, Manoel Limeira and Murta, Leonardo and Plastino, Alexandre},
  journal={Software: Practice and Experience},
  volume={51},
  number={6},
  pages={1173--1193},
  year={2021},
  publisher={Wiley Online Library}
}

@inproceedings{guo2024exploring,
  title={Exploring the potential of chatgpt in automated code refinement: An empirical study},
  author={Guo, Qi and Cao, Junming and Xie, Xiaofei and Liu, Shangqing and Li, Xiaohong and Chen, Bihuan and Peng, Xin},
  booktitle={Proceedings of the 46th IEEE/ACM International Conference on Software Engineering},
  pages={1--13},
  year={2024}
}

@inproceedings{grewal2024analyzing,
  title={Analyzing Developer Use of ChatGPT Generated Code in Open Source GitHub Projects},
  author={Grewal, Balreet and Lu, Wentao and Nadi, Sarah and Bezemer, Cor-Paul},
  booktitle={2024 IEEE/ACM 21st International Conference on Mining Software Repositories (MSR)},
  pages={157--161},
  year={2024},
  organization={IEEE}
}

@article{nakagawa2017coefficient,
  title={The coefficient of determination R 2 and intra-class correlation coefficient from generalized linear mixed-effects models revisited and expanded},
  author={Nakagawa, Shinichi and Johnson, Paul CD and Schielzeth, Holger},
  journal={Journal of the Royal Society Interface},
  volume={14},
  number={134},
  pages={20170213},
  year={2017},
  publisher={The Royal Society}
}

@article{chavalarias2016evolution,
  title={Evolution of reporting P values in the biomedical literature, 1990-2015},
  author={Chavalarias, David and Wallach, Joshua David and Li, Alvin Ho Ting and Ioannidis, John PA},
  journal={Jama},
  volume={315},
  number={11},
  pages={1141--1148},
  year={2016},
  publisher={American Medical Association}
}

@inproceedings{kudrjavets2022mining,
  title={Mining code review data to understand waiting times between acceptance and merging: An empirical analysis},
  author={Kudrjavets, Gunnar and Kumar, Aditya and Nagappan, Nachiappan and Rastogi, Ayushi},
  booktitle={Proceedings of the 19th International Conference on Mining Software Repositories},
  pages={579--590},
  year={2022}
}

@inproceedings{bernardo2018studying,
  title={Studying the impact of adopting continuous integration on the delivery time of pull requests},
  author={Bernardo, Jo{\~a}o Helis and da Costa, Daniel Alencar and Kulesza, Uir{\'a}},
  booktitle={Proceedings of the 15th International Conference on Mining Software Repositories},
  pages={131--141},
  year={2018}
}

@inproceedings{gousios2014exploratory,
  title={An exploratory study of the pull-based software development model},
  author={Gousios, Georgios and Pinzger, Martin and Deursen, Arie van},
  booktitle={Proceedings of the 36th international conference on software engineering},
  pages={345--355},
  year={2014}
}

@article{baysal2016investigating,
  title={Investigating technical and non-technical factors influencing modern code review},
  author={Baysal, Olga and Kononenko, Oleksii and Holmes, Reid and Godfrey, Michael W},
  journal={Empirical Software Engineering},
  volume={21},
  pages={932--959},
  year={2016},
  publisher={Springer}
}

@article{mann1947test,
  title={On a test of whether one of two random variables is stochastically larger than the other},
  author={Mann, Henry B and Whitney, Donald R},
  journal={The annals of mathematical statistics},
  pages={50--60},
  year={1947},
  publisher={JSTOR}
}

@article{nachar2008mann,
  title={The Mann-Whitney U: A test for assessing whether two independent samples come from the same distribution},
  author={Nachar, Nadim and others},
  journal={Tutorials in quantitative Methods for Psychology},
  volume={4},
  number={1},
  pages={13--20},
  year={2008}
}

@inproceedings{silva2020measuring,
  title={Measuring Unique Changes: How do Distinct Changes Affect the Size and Lifetime of Pull Requests?},
  author={Silva, Daniel Augusto Nunes da and Soares, Daric{\'e}lio Moreira and Gon{\c{c}}alves, Silvana Andrade},
  booktitle={Proceedings of the 14th Brazilian Symposium on Software Components, Architectures, and Reuse},
  pages={121--130},
  year={2020}
}

@article{JSSv094i12,
 title={PResiduals: An R Package for Residual Analysis Using Probability-Scale Residuals},
 volume={94},
 url={https://www.jstatsoft.org/index.php/jss/article/view/v094i12},
 doi={10.18637/jss.v094.i12},
 number={12},
 journal={Journal of Statistical Software},
 author={Liu, Qi and Shepherd, Bryan and Li, Chun},
 year={2020},
 pages={1–27}
}

@article{ye2021recommending,
  title={Recommending pull request reviewers based on code changes},
  author={Ye, Xin and Zheng, Yongjie and Aljedaani, Wajdi and Mkaouer, Mohamed Wiem},
  journal={Soft Computing},
  volume={25},
  pages={5619--5632},
  year={2021},
  publisher={Springer}
}

@inproceedings{malkauthekar2013analysis,
  title={Analysis of euclidean distance and manhattan distance measure in face recognition},
  author={Malkauthekar, MD},
  booktitle={Third International Conference on Computational Intelligence and Information Technology (CIIT 2013)},
  pages={503--507},
  year={2013},
  organization={IET}
}

@article{feng2014log,
  title={Log-transformation and its implications for data analysis.},
  author={Feng, Changyong and Wang, Hongyue and Lu, Naiji and Chen, Tian and He, Hua and Lu, Ying and Tu, Xin M},
  journal={Shanghai archives of psychiatry},
  volume={26},
  number={2},
  pages={105--109},
  year={2014}
}

@inproceedings{watanabe2024use,
  title={On the Use of ChatGPT for Code Review: Do Developers Like Reviews By ChatGPT?},
  author={Watanabe, Miku and Kashiwa, Yutaro and Lin, Bin and Hirao, Toshiki and Yamaguchi, Ken'Ichi and Iida, Hajimu},
  booktitle={Proceedings of the 28th International Conference on Evaluation and Assessment in Software Engineering},
  pages={375--380},
  year={2024}
}

@inproceedings{chouchen2024so,
  title={How Do Software Developers Use ChatGPT? An Exploratory Study on GitHub Pull Requests},
  author={Chouchen, Moataz and Bessghaier, Narjes and Begoug, Mahi and Ouni, Ali and AlOmar, Eman Abdullah and Mkaouer, Mohamed Wiem},
  booktitle={2024 IEEE/ACM 21st International Conference on Mining Software Repositories (MSR)},
  pages={212--216},
  year={2024},
  organization={IEEE}
}

@online{pullrequest2059,
  author       = {alshedivat},
  title        = {Pull Request \#2059},
  year         = {2018},
  url          = {https://github.com/alshedivat/al-folio/pull/2059},
  note         = {Accessed on: Oct. 28, 2024}
}

@online{pullrequest1858,
  author       = {qiyunzhu},
  title        = {Pull Request \#1858},
  year         = {2023},
  url          = {https://github.com/scikit-bio/scikit-bio/pull/1858#issuecomment-1583219957},
  note         = {Accessed on: Oct. 28, 2024}
}

@online{pullrequest188,
  author       = {JohnBeres },
  title        = {Pull Request \#188},
  year         = {2023},
  url          = {https://github.com/flaree/flare-cogs/pull/188},
  note         = {Accessed on: Oct. 28, 2024}
}

@online{pullrequest1171,
  author       = {haileyschoelkopf},
  title        = {Pull Request \#1171},
  year         = {2023},
  url          = {https://github.com/EleutherAI/lm-evaluation-harness/pull/1171},
  note         = {Accessed on: Oct. 28, 2024}
}

@online{pullrequest4616,
  author       = {alexander-cit},
  title        = {Pull Request \#4616},
  year         = {2023},
  url = {https://github.com/CitizenLabDotCo/citizenlab/pull/4616#issuecomment-1514994078},
  note         = {Accessed on: Oct. 28, 2024}
}

@online{pullrequest43,
  author       = {dciborow},
  title        = {Pull Request \#43},
  year         = {2023},
  url = {https://github.com/microsoft/gpt-review/pull/43/},
  note         = {Accessed on: Oct. 28, 2024}
}

@online{pullrequest238 ,
  author       = {islxyqwe},
  title        = {Pull Request \#238 },
  year         = {2023},
  url = {https://github.com/Kanaries/graphic-walker/pull/238},
  note         = {Accessed on: Oct. 28, 2024}
}

@online{pullrequest3108,
  author       = {trackleft},
  title        = {Pull Request \#3108},
  year         = {2024},
  url = {https://github.com/az-digital/az_quickstart/pull/3109#issuecomment-1901093471},
  note         = {Accessed on: Oct. 28, 2024}
}

@online{pullrequest36,
  author       = {cloud26},
  title        = {Pull Request \#36},
  year         = {2023},
  url = {https://github.com/codedog-ai/codedog/pull/36},
  note         = {Accessed on: Oct. 28, 2024}
}

@online{pullrequest736,
  author       = {okotek},
  title        = {Pull Request \#736},
  year         = {2024},
  url = {https://github.com/Codium-ai/pr-agent/pull/736},
  note         = {Accessed on: Oct. 28, 2024}
}

@online{pullrequest109,
  author       = {rudolfolah},
  title        = {Pull Request \#109},
  year         = {2023},
  url = {https://github.com/rhasspy/piper/pull/109},
  note         = {Accessed on: Oct. 28, 2024}
}

@online{pullrequest28300,
  author       = {potiuk},
  title        = {Pull Request \#28300},
  year         = {2022},
  url = {https://github.com/apache/airflow/pull/28300},
  note         = {Accessed on: Oct. 28, 2024}
}

@online{pullrequest27905,
  author       = {rusackas},
  title        = {Pull Request \#27905},
  year         = {2024},
  url = {https://github.com/apache/superset/pull/27905},
  note         = {Accessed on: Oct. 28, 2024}
}

@online{pullrequest1122,
  author       = {yceballost},
  title        = {Pull Request \#1122},
  year         = {2023},
  url = {https://github.com/Telefonica/mistica-design/pull/1122},
  note         = {Accessed on: Oct. 28, 2024}
}

@online{pullrequest31,
  author       = {clearlysid},
  title        = {Pull Request \#31},
  year         = {2024},
  url = {https://github.com/helmerapp/micro/pull/31},
  note         = {Accessed on: Oct. 28, 2024}
}

@online{GitHubTimeline,
  author       = {GitHub},
  title        = {REST API endpoints for timeline events},
  year         = {2022},
  url = {https://docs.github.com/en/rest/issues/timeline?apiVersion=2022-11-28},
  note         = {Accessed on: Oct. 28, 2024}
}

@online{pullrequest16,
  author       = {Rbx2Source},
  title        = {Pull Request \#16},
  year         = {2024},
  url = {https://github.com/LockpickInteractive/Rbx2Source/pull/16},
  note         = {Accessed on: Oct. 28, 2024}
}

@online{pullrequest1,
  author       = {meowstatus},
  title        = {Pull Request \#1},
  year         = {2024},
  url = {https://github.com/aludmilagdev/meow_status/pull/1},
  note         = {Accessed on: Oct. 28, 2024}
}

@online{pullrequest64,
  author       = {onasito},
  title        = {Pull Request \#64},
  year         = {2024},
  url = {https://github.com/onasito/Group-5/pull/64},
  note         = {Accessed on: Oct. 28, 2024}
}

@article{ebert2023generative,
  title={Generative AI for software practitioners},
  author={Ebert, Christof and Louridas, Panos},
  journal={IEEE Software},
  volume={40},
  number={4},
  pages={30--38},
  year={2023},
  publisher={IEEE}
}

@article{rasheed2024ai,
  title={Ai-powered code review with llms: Early results},
  author={Rasheed, Zeeshan and Sami, Malik Abdul and Waseem, Muhammad and Kemell, Kai-Kristian and Wang, Xiaofeng and Nguyen, Anh and Syst{\"a}, Kari and Abrahamsson, Pekka},
  journal={arXiv preprint arXiv:2404.18496},
  year={2024}
}

@inproceedings{vijayvergiya2024ai,
  title={Ai-assisted assessment of coding practices in modern code review},
  author={Vijayvergiya, Manushree and Salawa, Ma{\l}gorzata and Budiseli{\'c}, Ivan and Zheng, Dan and Lamblin, Pascal and Ivankovi{\'c}, Marko and Carin, Juanjo and Lewko, Mateusz and Andonov, Jovan and Petrovi{\'c}, Goran and others},
  booktitle={Proceedings of the 1st ACM International Conference on AI-Powered Software},
  pages={85--93},
  year={2024}
}

@article{davila2021systematic,
  title={A systematic literature review and taxonomy of modern code review},
  author={Davila, Nicole and Nunes, Ingrid},
  journal={Journal of Systems and Software},
  volume={177},
  pages={110951},
  year={2021},
  publisher={Elsevier}
}

@article{borges2018s,
  title={What’s in a github star? understanding repository starring practices in a social coding platform},
  author={Borges, Hudson and Valente, Marco Tulio},
  journal={Journal of Systems and Software},
  volume={146},
  pages={112--129},
  year={2018},
  publisher={Elsevier}
}

@article{vaswani2017attention,
  title={Attention is all you need},
  author={Vaswani, Ashish and Shazeer, Noam and Parmar, Niki and Uszkoreit, Jakob and Jones, Llion and Gomez, Aidan N and Kaiser, {\L}ukasz and Polosukhin, Illia},
  journal={Advances in neural information processing systems},
  volume={30},
  year={2017}
}

@inproceedings{sadowski2018modern,
  title={Modern code review: a case study at google},
  author={Sadowski, Caitlin and S{\"o}derberg, Emma and Church, Luke and Sipko, Michal and Bacchelli, Alberto},
  booktitle={Proceedings of the 40th international conference on software engineering: Software engineering in practice},
  pages={181--190},
  year={2018}
}

@book{james2013introduction,
  title={An introduction to statistical learning},
  author={James, Gareth and Witten, Daniela and Hastie, Trevor and Tibshirani, Robert and others},
  volume={112},
  pages={50--128},
  number={1},
  year={2013},
  publisher={Springer}
}

@inbook{jurafsky2023slp3,
  author    = {Daniel Jurafsky and James H. Martin},
  title     = {Speech and Language Processing (3rd ed. draft)},
  chapter   = {3: N-gram Language Models},
  year      = {2023},
  note      = {\url{https://web.stanford.edu/~jurafsky/slp3/3.pdf}},
  publisher = {Draft online version},
  howpublished = {Available online},
}

@article{zheng2024code,
  title={Code-Survey: An LLM-Driven Methodology for Analyzing Large-Scale Codebases},
  author={Zheng, Yusheng and Yang, Yiwei and Tu, Haoqin and Huang, Yuxi},
  journal={arXiv preprint arXiv:2410.01837},
  year={2024}
}

@article{cole2017statistics,
  title={Statistics Notes: What is a percentage difference?},
  author={Cole, Tim J and Altman, Douglas G},
  journal={Bmj},
  volume={358},
  year={2017},
  publisher={British Medical Journal Publishing Group}
}

@inproceedings{cihan2025automated,
  title={Automated code review in practice},
  author={Cihan, Umut and Haratian, Vahid and {\.I}{\c{c}}{\"o}z, Arda and G{\"u}l, Mert Kaan and Devran, {\"O}mercan and Bayendur, Emircan Furkan and U{\c{c}}ar, Baykal Mehmet and T{\"u}z{\"u}n, Eray},
  booktitle={2025 IEEE/ACM 47th International Conference on Software Engineering: Software Engineering in Practice (ICSE-SEIP)},
  pages={425--436},
  year={2025},
  organization={IEEE}
}

@article{ramesh2025automated,
  title={Automated Code Review Using Large Language Models at Ericsson: An Experience Report},
  author={Ramesh, Shweta and Bose, Joy and Singh, Hamender and Raghavan, AK and Roychowdhury, Sujoy and Sridhara, Giriprasad and Saini, Nishrith and Britto, Ricardo},
  journal={arXiv preprint arXiv:2507.19115},
  year={2025}
}

@book{krippendorff2011,
  title={Content Analysis: An Introduction to Its Methodology},
  author={Krippendorff, Klaus},
  year={2011},
  publisher={SAGE Publications}
}

@article{khatoonabadi2023wasted,
  title={On wasted contributions: Understanding the dynamics of contributor-abandoned pull requests---A mixed-methods study of 10 large open-source projects},
  author={Khatoonabadi, SayedHassan and Costa, Diego Elias and Abdalkareem, Rabe and Shihab, Emad},
  journal={ACM Transactions on Software Engineering and Methodology},
  volume={32},
  number={1},
  pages={1--39},
  year={2023},
  publisher={ACM New York, NY}
}

@article{khatoonabadi2024predicting,
  title={Predicting the first response latency of maintainers and contributors in pull requests},
  author={Khatoonabadi, SayedHassan and Abdellatif, Ahmad and Costa, Diego Elias and Shihab, Emad},
  journal={IEEE Transactions on Software Engineering},
  year={2024},
  publisher={IEEE}
}

@article{widyasari2023explaining,
  title={Explaining explanation: An empirical study on explanation in code reviews},
  author={Widyasari, Ratnadira and Zhang, Ting and Bouraffa, Abir and Maalej, Walid and Lo, David},
  journal={arXiv preprint arXiv:2311.09020},
  year={2023}
}

@inproceedings{bacchelli2013expectations,
  title={Expectations, outcomes, and challenges of modern code review},
  author={Bacchelli, Alberto and Bird, Christian},
  booktitle={2013 35th International Conference on Software Engineering (ICSE)},
  pages={712--721},
  year={2013},
  organization={IEEE}
}

@article{xiao2024generative,
  title={Generative AI for Pull Request Descriptions: Adoption, Impact, and Developer Interventions},
  author={Xiao, Tao and Hata, Hideaki and Treude, Christoph and Matsumoto, Kenichi},
  journal={Proceedings of the ACM on Software Engineering},
  volume={1},
  number={FSE},
  pages={1043--1065},
  year={2024},
  publisher={ACM New York, NY, USA}
}

@article{yu2024fine,
  title={Fine-tuning large language models to improve accuracy and comprehensibility of automated code review},
  author={Yu, Yongda and Rong, Guoping and Shen, Haifeng and Zhang, He and Shao, Dong and Wang, Min and Wei, Zhao and Xu, Yong and Wang, Juhong},
  journal={ACM Transactions on Software Engineering and Methodology},
  year={2024},
  publisher={ACM New York, NY}
}

@article{davila2024tales,
  title={Tales from the Trenches: Expectations and Challenges from Practice for Code Review in the Generative AI Era},
  author={Davila, Nicole and Melegati, Jorge and Wiese, Igor},
  journal={IEEE Software},
  year={2024},
  publisher={IEEE}
}

@article{hao2024empirical,
  title={An empirical study on developers’ shared conversations with ChatGPT in GitHub pull requests and issues},
  author={Hao, Huizi and Hasan, Kazi Amit and Qin, Hong and Macedo, Marcos and Tian, Yuan and Ding, Steven HH and Hassan, Ahmed E},
  journal={Empirical Software Engineering},
  volume={29},
  number={6},
  pages={150},
  year={2024},
  publisher={Springer}
}

@inproceedings{das2024investigating,
  title={Investigating the Utility of ChatGPT in the Issue Tracking System: An Exploratory Study},
  author={Das, Joy Krishan and Mondal, Saikat and Roy, Chanchal},
  booktitle={Proceedings of the 21st International Conference on Mining Software Repositories},
  pages={217--221},
  year={2024}
}

\appendix
\appendix
\section{Tables}
\label{sec:appendix-b}
In the following, we provide the GPT PRs during a Round 1, the accuracy across rounds after applying heuristics, and a summary of the GPT Assisted PR Dataset by Status.

\begin{table}[H]
\centering
\caption{GPT PRs during a Round of Sampling Analysis}
\label{tab:frequency_pattern_pr_r1}
\begin{tabular}{lrr}\hline
\hline
\textbf{Filter} & \textbf{Assistance} & \textbf{Non Assistance} \\ 
Project name & 4  & 21 \\ 
PR title & 6  & 19 \\ 
Modified files & 23 & 58 \\ 
PR body & 1  & 24 \\
\textbf{Total} & \textbf{34} & \textbf{122} \\ \hline
\end{tabular}
\begin{flushleft}
\footnotesize{Note: Some GPT-assisted PRs matched two or more filters simultaneously. The total count (34) reflects overlaps between categories, not distinct PRs per filter. For example, a single PR might be identified as GPT-assisted based on both its title and modified files, contributing to multiple row counts.}
\end{flushleft}
\end{table}

\begin{table}[H]
\centering
\caption{Accuracy Improvement across Rounds after Applying Heuristics}
\label{tab:frequency_pattern_pr_percetage}
\begin{tabular}{cccc}\hline
\hline
\textbf{Heuristic} & \textbf{Round 1 (\%)} & \textbf{Round 2 (\%)} & \textbf{Round 3 (\%)} \\ \hline 
H1 & 82 & 98 & 98 \\ 
H2 & 98 & 97 & 100 \\ 
H3 & 96 & 97 & 97 \\ 
H4 & 66 & 95 & 97 \\
\hline
\end{tabular}
\begin{flushleft}
\footnotesize{Note: Each round corresponds to one iteration of refinement, where heuristics were introduced to correct common misclassifications.}
\end{flushleft}
\end{table}

\begin{table}[H]
\centering
\caption{Summary: GPT Assisted PR Dataset by Status}
\begin{tabular}{lr}\hline
\hline
\textbf{Status} & \textbf{Assistance} \\
\hline 
Open & 54   \\ 
Closed & 179   \\ 
Merged & 1367 \\  
\textbf{Total} & \textbf{1600} \\ \hline
\end{tabular}
\label{tab:GPT-PR-Status-Summary}
\end{table}

\end{document}